\documentclass[twocolumn,trackchanges]{aastex631}

\begin{document}

\title{The CUISINES Framework for Conducting Exoplanet Model Intercomparison Projects, Version 1.0}

\author[0000-0002-6673-2007]{Linda E. Sohl}
\correspondingauthor{Linda E. Sohl}
\email{Linda.Sohl@columbia.edu}
\affiliation{Center for Climate Systems Research, Columbia University, 2880 Broadway, New York, NY, 10025, USA}
\affiliation{NASA Goddard Institute for Space Studies, 2880 Broadway, New York, NY, 10025, USA}
\affiliation{NASA GSFC Sellers Exoplanet Environments Collaboration, USA}

\author[0000-0002-5967-9631]{Thomas J. Fauchez}
\affiliation{NASA GSFC Sellers Exoplanet Environments Collaboration, USA}
\affiliation{NASA Goddard Space Flight Center, 8800 Greenbelt Road, Greenbelt, MD, 20771, USA}
\affiliation{American University, College of Arts and Science, Washington, DC, 20016 USA}

\author[0000-0003-0354-9325]{Shawn Domagal-Goldman}
\affiliation{NASA GSFC Sellers Exoplanet Environments Collaboration, USA}
\affiliation{NASA Goddard Space Flight Center, 8800 Greenbelt Road, Greenbelt, MD, 20771, USA}

\author[0000-0002-4997-0847]{Duncan A. Christie}
\affiliation{Max Planck Institute for Astronomy, K\"onigstuhl 17, D-69117 Heidelberg, Germany}
\affiliation{Physics and Astronomy, Faculty of Environment, Science and Economy, University of Exeter, Exeter, UK}

\author[0000-0001-9423-8121]{Russell Deitrick}
\affiliation{School of Earth and Ocean Sciences, University of Victoria, Victoria, BC, V8P 5C2, Canada}

\author[0000-0003-4346-2611]{Jacob Haqq-Misra}
\affiliation{Blue Marble Space Institute of Science, 600 1st Avenue, Seattle, WA, 98104, USA}

\author[0000-0003-2281-1990]{C. E. Harman}
\affiliation{NASA Ames Research Center, Moffet Field, CA, 94035, USA}

\author[0000-0003-2329-418X]{Nicolas Iro}
\affiliation{Institute of Planetary Research, German Aerospace Center (DLR), Rutherfordstrasse 2, D-12489 Berlin, Germany}

\author[0000-0001-6707-4563]{Nathan J. Mayne}
\affiliation{Physics and Astronomy, Faculty of Environment, Science and Economy, University of Exeter, Exeter, UK}

\author[0000-0001-5328-819X]{Kostas Tsigaridis}
\affiliation{Center for Climate Systems Research, Columbia University, 2880 Broadway, New York, NY, 10025, USA}
\affiliation{NASA Goddard Institute for Space Studies, 2880 Broadway, New York, NY, 10025, USA}
\affiliation{NASA GSFC Sellers Exoplanet Environments Collaboration, USA}

\author[0000-0002-2662-5776]{Geronimo L. Villanueva}
\affiliation{NASA GSFC Sellers Exoplanet Environments Collaboration, USA}
\affiliation{NASA Goddard Space Flight Center, 8800 Greenbelt Road, Greenbelt, MD, 20771, USA}

\author[0000-0003-3099-1506]{Amber V. Young}
\affiliation{NASA GSFC Sellers Exoplanet Environments Collaboration, USA}
\affiliation{NASA Goddard Space Flight Center, 8800 Greenbelt Road, Greenbelt, MD, 20771, USA}

\author[0000-0003-4711-3099]{Guillaume Chaverot}
\affiliation{Univ. Grenoble Alpes, CNRS, IPAG, 38000 Grenoble, France}

\begin{abstract}
As JWST begins to return observations, it is more important than ever that exoplanet climate models can consistently and correctly predict the observability of exoplanets, retrieval of their data, and interpretation of planetary environments from that data. Model intercomparisons play a crucial role in this context, especially now when few data are available to validate model predictions. The CUISINES Working Group of NASA’s Nexus for Exoplanet Systems Science (NExSS) supports a systematized approach to evaluating the performance of exoplanet models, and provides here a framework for conducting community-organized exoplanet Model Intercomparison Projects (exoMIPs). The CUISINES framework adapts Earth climate community practices specifically for the needs of the exoplanet researchers, encompassing a range of model types, planetary targets, and parameter space studies. It is intended to help researchers to work collectively, equitably, and openly toward common goals. The CUISINES framework rests on five principles: 1) Define in advance what research question(s) the exoMIP is intended to address. 2) Create an experimental design that maximizes community participation, and advertise it widely. 3) Plan a project timeline that allows all exoMIP members to participate fully. 4) Generate data products from model output for direct comparison to observations. 5) Create a data management plan that is workable in the present and scalable for the future. Within the first years of its existence, CUISINES is  already providing logistical support to 10 exoMIPs, and will continue to host annual workshops for further community feedback and presentation of new exoMIP ideas.
\end{abstract}

\section{Introduction} \label{sec:intro}

Since the first exoplanet orbiting a main sequence star was discovered \citep{Mayor_1995}, the existence of nearly 5,600 exoplanets has been confirmed, with over 7,500 additional exoplanets awaiting confirmation,\footnote{Per the NASA Exoplanet Archive as of March 2024, \\ \url{https://exoplanetarchive.ipac.caltech.edu/}} and the promise of many more discoveries to come from missions such as JWST and the future Habitable Exoplanet Observatory \citep{NAP26141}. It is clear even from basic demographics of bulk quantities such as mass and radius that exoplanets are extremely diverse. Not surprisingly, planetary scientists and astrobiologists are eager to explore the “climates” of these worlds – everything from determining the impacts of atmosphere compositions and stellar spectra on global mean temperatures to the potential for habitable surfaces on rocky planets, and to the intricacies of atmospheric (and oceanic) circulation on worlds of all kinds. A full range of atmosphere and related models is available: Radiative Transfer and Retrieval Models, 1-D Radiative-Convective Equilibrium Models (RCEs), Energy Balance Models (EBMs), Models of Intermediate Complexity (MICs), and 3-D General Circulation Models (GCMs). Some of these models were developed specifically for exoplanets. Others have been generalized for broader use from models developed for modern Earth; “recent” (from an astronomical perspective) paleo-Earth, i.e., oxygenated atmospheres of the past several $10^8$ years; and early Earth, marked by habitable conditions and anoxygenic atmospheres \citep{2001Natur.409.1083N}. With so many exoplanets to simulate, a plethora of models to choose from, and so few data to provide constraints, the planetary science community has been energized to run an ever-increasing number of experiments exploring the possible climates of particular planetary targets, or generalizable effects of stellar spectra and orbital parameters on simplified planetary surfaces such as aquaplanets, land planets, and “Snowball Earths.”  Early results have demonstrated the power of JWST observations \citep{Ahrer_2023,Alderson_2023,Feinstein_2023,Rustamkulov_2023}, placing the planetary community on the cusp of a potential step change in the quality and atmospheric data for a number of exoplanets, many different from anything in our Solar System, e.g. sub-Neptunes and Hot Jupiters. 

Despite all of the activity in adapting and applying models to exoplanets, it is often difficult to understand the extent to which an experiment result is a product of the model’s performance, or the initial or boundary conditions used for the experiment. Individual modeling groups commonly make modifications to model code and/or experiment configurations in order to successfully complete an experiment. However, those modifications are frequently considered technical and are not always fully described in publications, so any attempt at replicating a study with another model can be hampered by incomplete information. Moreover, planetary climate modelers as a community have not yet worked out standard ways of describing certain inputs such as modern Earth-like atmospheric composition for “control” experiments. So even when all experimental conditions are fully described, it is difficult to compare the results of separate studies that do not all have the same starting point. 

Model performance and skill will be crucial to consistently and correctly predicting the retrieval of the observational characteristics of these worlds, and the interpretation of planetary environments from that data. Collaborative activities such as model intercomparisons of identical experiments can promote improved understanding of the extent to which the results of participating models depend on numerical choices. This will be especially useful in the early stages of JWST’s deployment, when very few observation data are available to validate model predictions. 

The CUISINES\footnote{CUISINES: Climates Using Interactive Suites of Intercomparisons Nested for Exoplanet Studies. For more information, including links to projects, data, papers, and other updates, see \url{https://nexss.info/cuisines/}} Working Group of NASA’s Nexus for Exoplanet Systems Science (NExSS) Research Coordination Network has been established to develop and support a systematic approach to evaluating the performance of exoplanet models. In this paper, we describe a framework for conducting community-organized exoplanet Model Intercomparison Projects (exoMIPs) that is based upon similar longstanding efforts in the Earth climate science community, but modified for the particular concerns and needs of exoplanet climate modelers. We also introduce the first generation of exoMIPs officially supported by CUISINES.

\section{Model Intercomparison Projects as \\ Drivers of Model Development \\ and Community Engagement} \label{sec:2}

There have been previous efforts to compare results across exoplanet climate models and model types, providing examples of the types of information that can be gained. {\cite{Yang_2016}} compared differences in 1D radiative transfer calculations between two line-by-line codes (SMART and LBLRTM), a moderate resolution code (SBART), and four low-resolution codes that are used in GCMs (CAM3, CAM4$\_$Wolf, LMD-G, and AM2), simulating a planet with a modern Earth-like atmosphere and orbiting a G or M star. Small differences between the models were found when the surface temperature is lower than about 300 K. However, at higher temperatures, model predictions of radiative fluxes differed by tens of watts per square meter, mainly due to discrepancies in water vapor radiative transfer calculations, and primarily impacting the shortwave. The differences are also larger for an M-dwarf spectrum than a G-type spectrum. These results suggest that radiative transfer codes should be verified first before being used in an exoplanet GCM, especially for exoplanets near or beyond the inner edge of the habitable zone. Such exoplanets require a higher resolution of the near-IR H2O spectral absorption bands and windows than has typically been used before.

\cite{Polichtchouk_2014} is, to our knowledge, the first exoplanet GCM intercomparison, using five GCMs - BOB, CAM, ICGM, MITgcm, and PEQMOD - to study hot Jupiter atmospheres. All models solved the primitive equations, but used different numerical algorithms or grids. The key finding was that specific quantitative GCM predictions, such as the location of large vortices and hot spots, are strongly model dependent. A few years later, \cite{Yang_2019} initiated the first GCM intercomparison for a rocky exoplanet. They compared five GCMs - CAM3, CAM4, CAM4 Wolf, AM2, and LMDG (now Generic PCM, or G-PCM) - using simulations of both a rapidly rotating aqua planet receiving a G-star spectral energy distribution (SED) and a tidally locked aqua planet receiving an M-star SED. Relatively small differences (\textless 8 K) were found in global mean surface temperature predicted for cloudy exoplanets orbiting a G star, but large differences (20–30 K) were identified for cloudy planets orbiting M stars. These differences have been attributed to discrepancies in atmospheric dynamic, clouds, and radiative transfer. While clouds have been found to be the largest difference between the models, the interactions between radiative transfer (e.g. shortwave absorption by water vapor) and atmospheric circulation can also influence the atmospheric relative humidity and therefore affect the surface temperature.

We note that both of the above studies involved members of different modeling groups; the need to coordinate both efforts and the time invested in the collaboration likely contributed to their successful outcomes. These studies stand in contrast to another proposed exoplanet model intercomparison, the Palaeoclimate and Terrestrial Exoplanet Radiative Transfer Model Intercomparison Project \citep[PALAEOTRIP][]{Goldblatt_2017}), which was suggested by members of one modeling group. The objective of PALAEOTRIP was to compare a large variety of radiation codes used for paleoclimate or exoplanet sciences, and to identify the limit conditions for which each model can produce accurate results. Such an intercomparison would have been extremely useful; however, to our knowledge, no results have been published from that intercomparison. It appears that the call to participate in PALAEOTRIP did not reach a sufficiently wide audience of potential participants, and without buy-in from collaborators, the project was not able to proceed.

\subsection{The Origin of CUISINES}\label{sec21}

At the 2017 Habitable Worlds conference in Wyoming, a group of planetary scientists began to work out a plan for a GCM model intercomparison of climate experiments investigating TRAPPIST-1e, a prime candidate for observation and atmospheric characterization of a rocky exoplanet in the habitable zone. This effort became known as the TRAPPIST Habitable Atmosphere Intercomparison project, or THAI \citep{Fauchez_2020}. The THAI project culminated in a workshop held 2020 September 14-16 \citep{Fauchez_2021}, and several group papers have documented the results of the experiments across the participating models (\citealp{Turbet_2022,Sergeev_2022,Fauchez_2022}; and related papers\footnote{THAI focus issue of The Planetary Science Journal, \\  \url{https://iopscience.iop.org/collections/2632-3338_focus_issue_THAI/}}). The enthusiasm generated by the broad community participation in THAI was a strong indication that additional exoplanet model intercomparison projects could be viable.

As a result, the NASA Nexus for Exoplanet System Science (NExSS) supported the formation of a working group, Climates Using Interactive Suites of Intercomparisons Nested for Exoplanet Studies, or CUISINES. CUISINES has, as one of its key goals, the expansion and formalization of an exoplanet model intercomparison framework for additional projects encompassing all model types. NExSS sponsored the first CUISINES workshop in 2021 September 27-29, called BUFFET (Building a Unified Framework for Exoplanet Treatments), where participants discussed the adoption of a CUISINES framework based on the Earth science climate community’s long-lived Coupled Model Intercomparison Project (CMIP) and Paleoclimate Model Intercomparison Project (PMIP), but designed specifically to fit the needs of the planetary climates community.

\begin{deluxetable*}{lccc}
\tablecaption{The inaugural group of CUISINES exoMIPs. Extended details about each exoMIP can be found in the listed protocol papers. The exoMIP leads (Chefs) correspond to the first author of each protocol reference.}
\tablenum{1}
\tablehead{\colhead{ExoMIP name} & \colhead{Model type(s) used} & \colhead{Target/Purpose} & \colhead{Protocols reference}} 
\startdata
CAMEMBERT & 3-D GCMs & Mini-Neptunes (GJ 1214b and K2-18b) & \citet{Christie_2022} \\
CREME & 3-D GCMs, MICs & Earth viewed as an exoplanet & Tsigaridis et al. in prep. \\
COD-ACCRA & 1-D RCEs & Broad 1-D model comparisons via select experiments & Chaverot et al., TBD \\
FILLET & EBMs & Parameter space study for temperate Earth-like exoplanets & \citet{https://doi.org/10.48550/arxiv.2302.04980} \\
MALBEC & Radiative transfer codes & Broad RT model comparisons via select experiments & \citet{Villanueva_2024} \\
MOCHA & 3-D GCMs & Hot Jupiters & Iro et al. in prep. \\
PIE & 1-D Photochemistry models & Broad 1-D model comparisons via select experiments & Harman et al. in prep. \\
RISOTTO & Retrieval codes & Transit and direct imaging targets & Young et al., TBD \\
SAMOSA & 1-D RCEs, EBMs, 3-D GCMs & Parameter space study for planets orbiting M-stars & \citet{Haqq_Misra_2022} \\
THAI & 3-D GCMs & TRAPPIST-1e & \citet{Fauchez_2020} \\
\enddata
\end{deluxetable*}

Participants of BUFFET-1 also decided which new exoMIPs to do first (see Table 1). Because sub-Neptune exoplanets were expected to be thoroughly observed with JWST, CAMEMBERT was proposed, which utilizes GCM simulations of sub-Neptune atmospheres for GJ1214~b and K2-18b, two prime JWST cycle 1 targets. CAMEMBERT test cases are designed to separately evaluate the differences due to the dynamical core or the radiative transfer scheme. CREME is motivated by the need to benchmark exoplanet GCMs predictions of Earth with Earth observing data. FILLET concerns energy balance models (EBMs) which are widely used in the exoplanet community to predict ice distribution on exoplanets. EBMs employ a large variety of parameterizations that can significantly differ from an EBM to another, therefore leading to model dependencies. MALBEC provides a comparison of exoplanet spectrum generators for which parameterizations, linelist choice, etc., can significantly impact the model spectra. PIE is the exoMIP focused on 1-D photochemical models, a category of model largely used in the exoplanet community to simulate atmospheric composition around different kinds of stars. Radiative transfer considerations and chemical network differences could lead to different atmospheric predictions between these models that need to be assessed. SAMOSA aims to simulate sparse samples for a synchronously rotating planet within a large grid of surface pressure and instellation to recover the full parameter space using interpolations. Discrepancies in climate predictions at the sample points, due to intrinsic model differences, can change the heat map of the parameter space. While being principally for GCMs, SAMOSA is also open to lower complexity models (1D EBMs and radiative, convective and equilibrium (RCE) models). 

At the BUFFET-2 workshop held one year later (2022 October 20-21), these first CUISINES exoMIPs reported on their experiences and challenges in developing workable protocols for the diverse model types and research questions at the heart of each project. At this workshop, another item was added to the CUISINES menu: MOCHA, which focuses on assessing differences in the GCM dynamical core of hot and ultra hot Jupiters. The two most recent additions to CUISINES were announced during or shortly after the BUFFET-3 workshop (2023 October 10-11). COD-ACCRA is focused on 1-D RCEs, which have been used in countless exoplanet studies, and will use a similar target list as PIE. RISOTTO is the first exoMIP that focuses on retrieval codes for both transit and direct imaging targets.

Given the notable progress and community interest generated by these activities, CUISINES will continue to hold annual BUFFET workshops in the future, providing a community forum for interim reports on CUISINES project progress, presentation of new exoMIP ideas, and adaptation of new standards as the field continues to evolve.

\subsection{Inspiration for the CUISINES Framework: \\ CMIP and PMIP}\label{sec22}

To gain some perspective on what might be accomplished via community-driven exoMIPs, we look to Earth climate science and the Coupled Model Intercomparison Project (CMIP). CMIP today is a monumental worldwide endeavor.\footnote{WCRP Coupled Model Intercomparison Project (CMIP), \\  \url{https://www.wcrp-climate.org/wgcm-cmip/}} It combines the efforts of hundreds of researchers in 48 modeling groups using atmosphere-ocean GCMs, Earth system models (GCMs with biogeochemical modeling capabilities), and Earth models of Intermediate Complexity (EMICs) to conduct sets of coordinated multi-model experiment intercomparisons of past, present, and future climate scenarios, as well as smaller, specialized MIPs on selected topics of interest to the community.\footnote{See “Overview of all CMIP6-Endorsed Projects,” \\ \url{https://wcrp-cmip.org/mips/cmip6-endorsed-mips/}} The outcomes of CMIP’s work are most commonly associated with future climate change projections in the assessment reports published by the Intergovernmental Panel on Climate Change \citep[IPCC; see e.g.][]{Lee_future}, but CMIP also plays an important role in providing insights to modeling groups on model performance and development needs. 

Though it is now a global-scale effort, CMIP is an outgrowth from much humbler beginnings in the late 1980s and early 1990s, when the earliest intercomparisons were more of an ad hoc affair \citep{https://doi.org/10.1002/wcc.648}. At that time, GCMs and the necessary supercomputing infrastructure to run 3-D experiments were just common enough that roughly 10 modeling groups worldwide had the ability to run simple atmosphere-only climate experiments, testing parameterizations in the then-cutting edge climate models for use in future climate predictions. Not all modeling groups conducted all the same experiments, but what analyses and intercomparisons could be done were incorporated into the first IPCC Assessment Report \citep{cubasch1990processes}. 

Because of the strong interest in further model intercomparisons, CMIP was formally established as an endeavor of the World Climate Research Programme in 1995 \citep{Meehl_1997}. In its early phases, the experiments were designed to document systematic simulation errors in global climate models; understand why the errors occurred; find ways to fix the errors; and only then assess model performance in reproducing key aspects of Earth’s climate system \citep{Meehl_1997,WOS:000166407500001}. Years later, one of the principal goals of CMIP remains focused on promoting better understanding of model results, and enabling modeling groups to learn from each other \citep{Eyring_2016}. The literature is replete with papers that assess model skill and identify biases, especially between CMIP phases \citep[e.g.][]{2019JD032321,EvaluationofLeadingModesofClimateVariabilityintheCMIPArchives,Seferian2020,BenchmarkingPerformanceChangesintheSimulationofExtratropicalModesofVariabilityacrossCMIPGenerations,10.3389/fmars.2023.1168385,WOS:000886860400002}. Such assessments of climate model skill, unlike numerical weather prediction skill, are retrospective; they may utilize hindcasts of simulated mean climate states compared against past observations (instrument-measured or proxy-derived), or compare past climate projections against observed outcomes \citep[e.g.][]{https://doi.org/10.1002/wcc.58,cp-9-811-2013,https://doi.org/10.1029/2019GL085378}.

As GCMs have advanced in terms of capabilities and supercomputing resources have grown to accommodate them, both the number of experiments and the modeling groups engaged have increased, and the science questions have evolved in sophistication and specification. Not all modeling groups are interested in and/or are able to participate in every experiment that could be done. CMIP has therefore adopted an operational structure that consists of a small number of mandatory core experiments \citep[a.k.a., the "DECK"; see][]{Eyring_2016} that are simple enough to be performed by any model participating in CMIP. Ancillary specialized MIP projects, such as those focused on the oceans \citep{gmd-9-3231-2016}, carbon cycle modeling \citep{gmd-9-2853-2016}, and ice sheet modeling \citep{gmd-9-4521-2016}, have their own supplemental experimental designs and remain open to any interested and capable parties.

The CMIP endeavor is supported by a large data infrastructure through which experiment results from the different models are converted to a common file format and naming convention for the diagnostic variables, and then shared broadly so that all data users, regardless of affiliation, can readily access and compare the results for a variety of analyses and applications. The usefulness of these data for the assessment reports published by the Intergovernmental Panel on Climate Change \citep[IPCC; see e.g.][]{Lee_future} has led to a robust link between the two efforts. Indeed, as of the most recent CMIP \citep[CMIP6;][]{Eyring_2016}, the timeline for completing experiments and analyses for the intercomparisons was aligned to feed smoothly into the corresponding IPCC compendium (AR6) on the physical basis for climate change \citep{RN1}, an arrangement that will be kept for future CMIPs and IPCC assessments.

The Paleoclimate Modelling Intercomparison Project (PMIP)\footnote{Paleoclimate Modelling Intercomparison Project (PMIP), \\ \url{https://pmip.lsce.ipsl.fr/}} promotes MIPs focused on specific times in Earth history that have special interest not only for reconstructing past climates, but also for examining processes and feedback responses that may offer insights into how future climates might behave; some of these MIPs are also part of CMIP \citep{gmd-11-1033-2018}. The individual MIPs under the PMIP umbrella all develop their own protocols and schedules (\citealp[see e.g.][]{Haywood_2016,gmd-10-4005-2017,gmd-10-4035-2017,Lunt_2017,ot07100r}). This approach is necessary because the various time periods explored have different relationships to processes and feedbacks of interest, and the experiments themselves require different initial and boundary conditions. A fundamental part of these MIPs is the comparison of model results to paleoproxy data, which of course are not available for future Earth climate experiments. 

PMIP projects may also require some flexibility in approach: participants in a PMIP project may be using a wider variety of models compared to CMIP, including older generation models that do not have the most current capabilities. There may also not be sufficient resources to develop time-specific geographic boundary conditions for a given model, in which case those users would have to use modern Earth geography as a substitute instead. Lastly, five PMIP projects contribute only a single reference experiment to the CMIP data archive (e.g., past 1000 years, mid-Holocene, Last Glacial Maximum, Last Interglacial 127k, Mid-Pliocene). Results from additional experiments for these five projects, and PMIP projects that fall outside the joint scope of CMIP-PMIP (e.g., deglaciation experiments, \citealp{gmd-9-2563-2016,gmd-12-3649-2019} and DeepMIP \citep{Lunt_2017}; see \citet{gmd-11-1033-2018}) do not have the extensive support that CMIP has – especially for data storage and sharing, which is handled in whatever manner each MIP can arrange. 

Neither CMIP nor PMIP offer a ready blueprint for exoMIP planning and support. The exoplanet modeling community is both smaller (currently, perhaps 200 people at most) and more diverse (with respect to model types that span 1-D to 3-D) than the CMIP community. Study cases may also include planetary types ranging from temperate Earths to Hot Jupiters, in which the composition, physical processes, chemistry, etc. are not necessarily similar. At the same time, individual exoMIPs will often be interested in using different types of models to simulate climates of the same targets, using the same observations to analyze synthetic spectra generated from the modeled climates. This common focus will inevitably result in a need to support a greater degree of interaction between exoMIPs than is typically the case in the PMIP community, where each MIP can be conducted separately and without reference to any other project. A different approach is needed for the exoplanet modeling community at the outset: one that accommodates the needs and constraints of the planetary science field as they exist now, and can continue to do so as this field matures and more data are available for model performance assessments. 

\section{The CUISINES Framework: \\ Five Principles for exoMIP Design} \label{sec:3}

The following framework developed from discussions held at the THAI and BUFFET workshops, as well as during regular meetings of the individual exoMIPs. There are many considerations when constructing/designing a MIP; here we simply aim to prescribe a consistent framework enabling all exoMIPs to contribute to our wider understanding of what is needed to progress our modelling of exoplanets. ExoMIP planners who would like their project to be endorsed and promoted by CUISINES should utilize this framework and associated guidance in planning their projects. Some of the guidance is specific to group intercomparisons. However, all planetary climate modelers, working individually or collectively, are encouraged to adopt as many of these best practices as possible for the benefit of the community as a whole, especially with respect to the data management practices. 

The CUISINES framework adapts aspects of both CMIP (hierarchical project design; centralized information exchange, data management and metadata protocols) and PMIP (flexibility in model participation, comparison of model output to data for validation) into an overall framework suitable for any exoplanet model type. Outlined here are general principles that any exoMIP within CUISINES should follow:

\begin{itemize}
  \item Define in advance what research question(s) the exoMIP is intended to address.
  \item Create an experimental design that maximizes community participation, and advertise it widely.
  \item Plan a project timeline that allows all exoMIP members to participate fully.
  \item Generate data products from model output for comparison to observations using standardized atmosphere-to-spectra formats.
  \item Create a data management plan that is workable in the present and scalable for the future.
\end{itemize}

Individual exoMIPs will always require their own set of experimental protocols. However, the CUISINES framework principles are agnostic as to model type, exoplanet target, or intercomparison rationale, so anyone can utilize them. In the following subsections, each of these principles is discussed in greater depth.

\subsection{Define in advance what research question(s) the exoMIP is intended to address} \label{sec:31}

With the number of exoplanets to explore growing steadily, it might be tempting to start up an exoMIP whenever a new and exciting discovery is made, especially for those model types that can complete experiments very quickly. Working rapidly through many simulations of different targets does not seem advisable, though: BUFFET workshop participants have expressed the concern that rapid expansion of exoMIP projects may create the risk of “burning out” by constant churning through new experiments, none of which are receiving the attention they deserve (for example, comparison with previous studies of similar but not identical targets). For models like GCMs that take longer to complete experiments, the pace of work may be less hurried but there is a greater chance of discovering, some months down the line, that the simulations completed did not address a key point because they were not planned adequately.

To avoid such concerns, begin by asking questions such as: What science question can modeling \emph{this} particular target do to build upon of past knowledge and advance our understanding of similar targets more broadly? Does this target provide an opportunity to compare a model experiment directly with observations? Can this intercomparison resolve known inconsistencies in the literature between model results for a given target, when it is unknown whether model diversity or free parameters (or perhaps both!) have led to differing results? More generally: What model capabilities could be improved through specific tests of model performance? 

Exoplanet climate modelers can take a cue here from the Earth climate science community. Under CMIP, research topics of broad community interest are defined by the WCRP’s Grand Challenges \footnote{WRCP Grand Challenges, \url{https://www.wcrp-climate.org/grand-challenges/grand-challenges-overview}}. These topics have been developed through community input, and define both the most pressing needs for advancing the field in a meaningful way, as well as the most significant barriers to be overcome in resolving those needs. The Grand Challenges also define metrics for knowing when research goals have been reached. Lastly, the Grand Challenges provide storylines that engage the public, attract future talent, and improve interdisciplinary connections. 

There are similar documents in the planetary sciences realm that describe community-based areas of interest. The \textit{NASA Astrobiology Strategy 2015} \citep{NASA_Astrobio}, the \textit{AstRoMap European Astrobiology Roadmap} \citep{Horneck_2016}, the \textit{2023-2032 Decadal Strategy for Planetary Science and Astrobiology} \citep{NASEM_2022a}, and the \textit{Independent Review of the Community Report from the Biosignature Standards of Evidence Workshop} \citep{NASEM_2022b} all include key topics such as life and habitable environments in the Solar System; the potential for extraterrestrial life and observable biosignatures; and climate evolution on solid bodies. Many of these topics are addressable through well-planned planetary climate experiments, so consider whether a new target of interest might also be useful for tackling one of these topics. Furthermore, linking an exoMIP purpose to broad themes of community-wide interest creates interdisciplinary connections from modelers to colleagues whose focus is on field campaigns, robotic missions, or remote observations.

This principle is of course useful for anyone interested in starting an exoMIP. Potential exoMIP leaders who specifically want their project idea to be endorsed by CUISINES should contact CUISINES leadership for assistance in identifying possible collaborators and ensuring that the project will meet the CUISINES framework requirements. Opportunities for cross-exoMIP interactions are enabled by all contributors aligning with the framework outline in this paper. 

\subsection{Create an experimental design that maximizes community participation, and advertise it widely} \label{sec:32}

For the work to have greater significance to the community, it is important that each exoMIP include as many interested modeling groups and participants as possible. There are three considerations for this part of the framework.

First, encourage broad participation with a MIP experimental design similar to that of CMIP and various PMIP projects: Create a low barrier for entry by defining core experiment(s) that are intentionally simple in design (e.g., single component changes), so that every participating group can complete them (Figure 1). These experiments can also serve as benchmarks or controls for model performance in “known” scenarios like modern Earth. All core experiments should require minimal effort to set up and run, but should also be informative enough to be able to answer science questions, and not serve only as a technical backbone of limited scientific interest.

\begin{figure}[htb]
\plotone{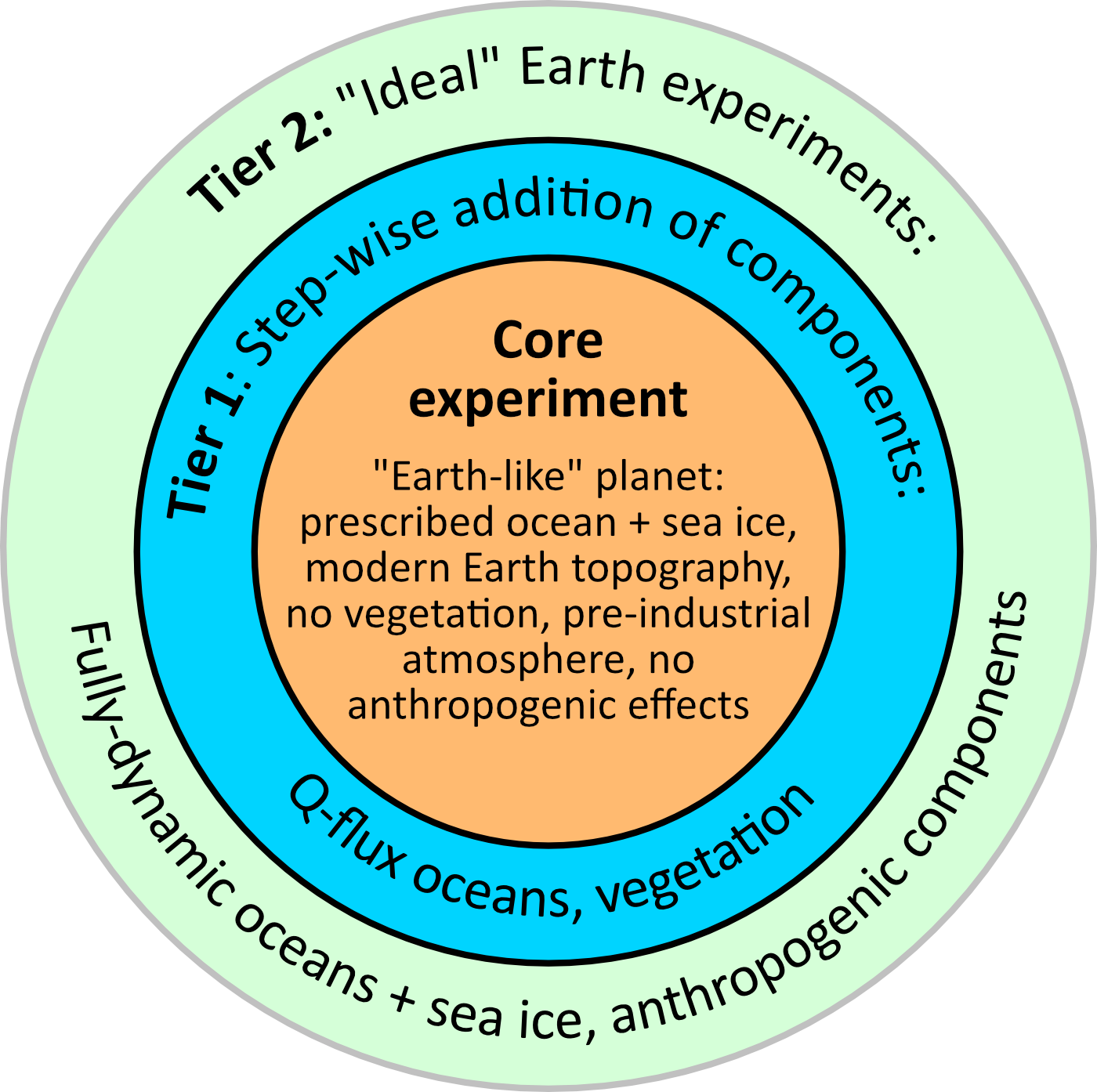}
\caption{Illustration of the general experimental design for the CREME exoMIP (Tsigaridis et al. in prep), as an example of how to structure an exoMIP to allow broad community participation.  
\label{fig:1}}
\end{figure}

Next, consider that not all modeling groups will have the personnel and/or technical resources to conduct experiments that require more effort, such as extended parameter space, complex scenarios, or extremely long simulations. Progressively more complex experiments and/or specialized experiments should be reserved for later phases of the MIP and be made optional, so that participating groups can complete them as resources and personnel availability permit. Similarly, experiments requiring additional model capabilities and/or extensive changes to boundary conditions can also be an obstacle to participation if personnel with the necessary technical expertise are not available. Part of the exoMIP planning process should consider how many groups are able to access that level of support. If there are not at least two or three capable modeling groups, reconsider whether such specialized experiments are truly informative for the exoMIP, then either postpone them or take them out of the project design altogether.

Lastly, once a tentative experimental design has been developed, the exoMIP leaders should announce it widely to attract additional participants, and solicit community feedback. Announcements via dedicated community email lists and social media, and presentations at conferences, are among the best ways to publicize the new project and to ensure that as many people as possible are given the opportunity to express interest. Note that all CUISINES-endorsed exoMIPs will receive assistance in advertising to interested parties, as well as in creating and managing a presence on the CUISINES website.

\subsection{Plan a project timeline that allows all exoMIP members to participate fully} \label{sec:33}

ExoMIP planners should define a timeline for designing and completing experiments, and contributing to group manuscripts, that reasonably accommodate the schedules of participating groups. The exoMIP will likely be an unfunded activity for many, and as such will not be a top priority. An activity like a MIP needs to fit around existing projects, course schedules, and other professional activities. Schedule flexibility also helps when (not if) the first efforts at simulating novel planetary environments are not successful, owing to technical obstacles.

We recommend scheduling a planning session for potential participants, in which the goals of the exoMIP are defined and a set of protocols drafted, taking into account the diverse specific needs of participating models (and their modelers). The annual BUFFET workshops will provide a venue for these planning conversations, though prospective participants in an exoMIP need not wait if they are ready to move forward. Planning sessions as side meetings at conferences can be helpful, but project leaders should also consider at least one virtual planning session so that remote participants may also contribute to the discussion.

With these tasks done, it becomes easier to estimate a schedule for experiment completion, the drafting of papers for individual model descriptions and/or results (as needed), and the preparation of the main group intercomparison paper for the exoMIP. Publishing the protocols early also helps to broaden participation by giving the community yet another opportunity to consider joining the project. Each MIP should establish a set of project milestones, to help communicate goals and progress. 

Researchers with limited time availability can become most deeply involved with the core experiments of an exoMIP, which should be the least difficult to complete successfully. A benefit to this approach is that these core experiments can become “benchmark cases” written up independently of the schedule of a full-scale exoMIP analysis, and still usefully contribute to the community’s body of knowledge. Indeed, CUISINES plans to collate benchmark cases from across all exoMIP efforts into a special collection called BASIL (Benchmark Atmospheric Simulations for Intercomparison Linkages), which can be used by researchers both within and outside of CUISINES to assess model performance for selected exoplanets and exoplanet types, as well as track the evolution of models (new models, or older models with new capabilities) over time. 

\subsection{Generate data products from model output for comparison to observations using standardized atmosphere-to-spectra formats} \label{sec:34}

Exoplanet model intercomparisons are important not only for highlighting similarities and differences between models, but in also their associated synthetic observations - which has ramifications for how future observations may be interpreted. For this reason, CUISINES-endorsed exoMIP projects must require that their participants link their model output to potentially observable characteristics through a synthetic spectrum generator (e.g., The Planetary Spectrum Generator (PSG\textbf{)}; \citealp{2022fpsg.book.....V}) and/or an instrument noise model, thermal phase curves, or albedo profiles. THAI illustrated the usefulness of this step, noting how comparison of GCM output discrepancies led to different predicted exoplanet spectra and therefore differing amounts of observation time to stipulate in JWST proposals \citep{Fauchez_2022}. 

\subsection{Create a data management plan that is workable in the present and scalable for the future} \label{sec:35}

Any given exoMIP project will likely not have the resources to support long-term data archiving and management, and public digital archives are not always suitable for exoplanet model output. At the same time, the adoption of open data and open science practices, such as those of the FAIR Guiding Principles,\footnote{FAIR Guiding Principles for Scientific Data Management and Stewardship, \url{https://www.go-fair.org/fair-principles/}} and NASA’s Open Source Science Initiative and associated Science Information Policy,\footnote{NASA Science Information Policy, \url{https://science.nasa.gov/researchers/science-data/science-information-policy}} makes data preservation and availability for decades a requirement. To \textbf{help }address this issue, CUISINES has prepared a platform for sharing output from CUISINES-endorsed exoMIPs that will support long-term data storage, and satisfies security protocols required by participating institutions. CUISINES employs the Comprehensive Knowledge Archive Network (CKAN) to store model data\footnote{Comprehensive Knowledge Archive Network, \url{https://ckan.emac.gsfc.nasa.gov/organization}} and GitHub to store scripts and input files.\footnote{CUISINES GitHub, \url{https://github.com/projectcuisines}}. In addition to the CKAN archive, exoMIP participants are welcome to link specific model data files to the published papers that utilize them, as well as draw attention to data sets that may be useful for downstream studies by publishing them in a data journal. Links to papers and data journal publications can easily be included in the list of products that CUISINES maintains for each exoMIP.

Because storing large quantities of model output is not necessary for the goals of an exoMIP, and may be intimidating to non-modelers interested in doing their own analyses of the experiment results, exoMIP planners should also consider:

\begin{itemize}
  \item The diagnostic variables needed for the immediate purposes of the exoMIP, and a separate list of those that are not needed urgently but are common enough to be useful for additional analyses later.
  \item The volume of model output to be made available, and the question of saving diagnostics as time-averaged climatologies, as time series, or both. This is an issue primarily for EBMs and GCMs, given the volume of raw output these models produce. 
  \item The conversion of file formats to widely used open formats, and ensuring that files are “future-proofed” to the greatest extent possible. 2-D and 3-D climate diagnostics should follow the NetCDF Climate and Forecast (CF) Metadata Convention. 1-D diagnostics and other necessary information, such as simulation configuration files or processing templates, should be in a human-readable form such as plain text files. No in-house or proprietary or commercial file formats should be used. 
  
  As an illustration, \cite{Villanueva_2024} have developed a “MALBEC.txt” file for the MALBEC exoMIP (Figure 2) that describes each specific case and then provides the necessary model input data for generating simulated spectra. A general python script converts the MALBEC.txt file into the input format required by any of the MALBEC participating models. Thus this file not only includes a unique format linking exoplanet atmospheric models and exoplanet radiative transfer models for each case, it provides documentation of the process as well. CUISINES will use this “MALBEC.txt” file format more generally to systematically connect the atmospheric outputs of each model type (with the exception of EBMs, which do not have an atmosphere, and of retrieval models which ingest and do not produce spectra) to synthetic spectra in a consistent and easily repeatable fashion.

\begin{figure*}[ht!]
\plotone{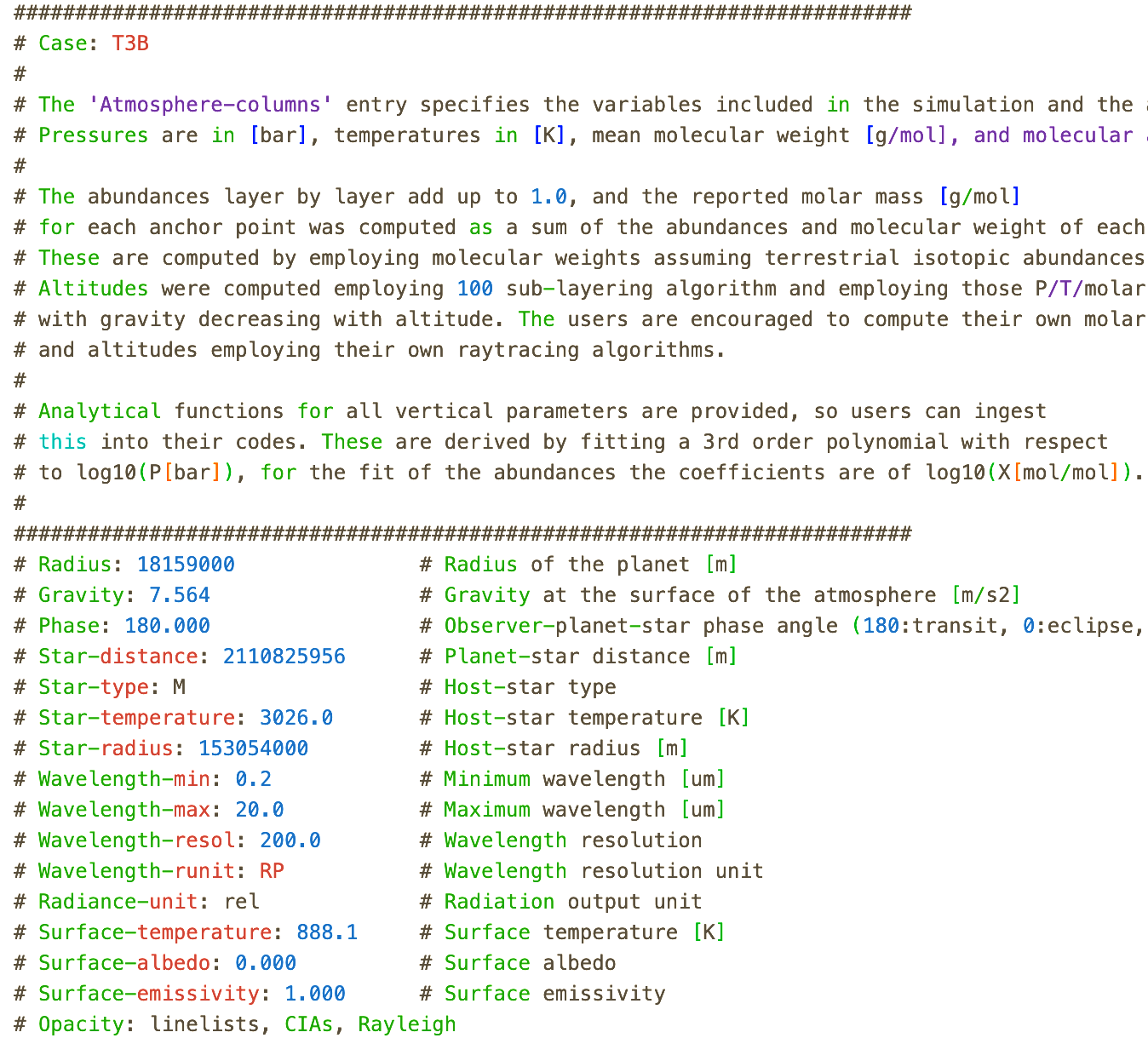}
\caption{Illustration of a ``MALBEC.txt" file created by \textbf{\cite{Villanueva_2024} }to generate synthetic spectra from model output. The file begins with a description of how the spectra are generated, followed by a list of the input variables and their values for this specific case. 
\label{fig:2}}
\end{figure*}
  
  \item The metadata to be provided by each modeling group such that the exoMIP name, experiment names, diagnostic variable definitions, and group contacts are clearly identified. Since many modeling groups use in-house climate diagnostic names and abbreviations, it will likely be necessary to provide a translation of in-house terminology to the CF Convention for standardized names and abbreviations already in use by the Earth climate science community. The CF Convention also provides guidelines for creating new diagnostic names as needed, so exoplanet modelers are not restricted to Earth-oriented diagnostics. Ideally these translations would be noted both within the output file metadata headers and, for the full set of output files contributed by a given modeling group, compiled in a separate plain text file for easy reference. 
\end{itemize}

Whenever data are shared, it is important to keep in mind that making data available (putting it in an online archive) is not the same as making it accessible (easily utilized by downstream users). If an objective is to entice research colleagues who are not climate modelers and other potentially interested parties (e.g., educators) into exploring the exoplanet climate modeling realm, the output should be made available without any requirement for additional post-processing. Raw model output and post-processing scripts should never be the default for data archive products.

With post-processed model output readily available via the CUISINES CKAN archive, other researchers have already begun to utilize THAI products. For example, models not currently represented in a CUISINES exoMIP have been benchmarked against the THAI output \citep{2023GMD....16.5601S,2024PSJ.....5....2R}. We anticipate that the results of other CUISINES exoMIPs will find similar applications in the future.

\section{Applying the CUISINES Framework Across Diverse Model Types} \label{sec:4}

Given the array of model types that will operate under the banner of CUISINES, individual exoMIPs will need to draft their own experiment protocols, since one size does not fit all. Each protocol should be developed with the five general principles in mind, but it also needs to be clear about what models and capabilities are specifically needed to join in a particular project, and the level of expertise and commitment that is required to participate fully in both core and optional experiments.

Eventually there may be targets of interest to multiple model types of varying complexity, resulting in a cross-model exoMIP for such targets and an opportunity to gain insights that might otherwise be missed. \cite{Maher_2019} described how high and low complexity models bring complementary information to understanding atmospheric physics; specifically, they demonstrated that a hierarchy of idealistic models is key in understanding complex systems. The large variety of models within the CUISINES framework will bring such similar complementarity, specifically when inputs/outputs will be connected between the exoMIPs.

A challenge for cross-model exoMIPs can arise in planning a schedule, since GCMs will always be significantly slower than 1-D models. A schedule with rapid completion times will always be in conflict with the longer time frame needed to complete GCM runs (typically weeks to months, compared to a few hours for 1-D models). The disparity has ramifications for the speed of completing intercomparison analyses and submitting manuscripts. This is not an insurmountable obstacle to cross-model interactions, however. The exoMIPs can be staged such that the faster models first spot-check a variety of scenarios, to identify the most interesting ones that would benefit from more in-depth explorations with EBMs and GCMs. Once the output from those EBM/GCM runs is available, their output could be used to identify adjustments that would be beneficial for the faster models or used as inputs for specific target scenarios. For example, the MALBEC exoMIP for radiative transfer codes \citep{Villanueva_2024} is starting with generic inputs for the first version of the intercomparison; once outputs from the CAMEMBERT, CREME and PIE exo-MIPs are available, a second version will be performed. 

Similarly, 1-D model exoMIPs such as PIE would benefit from coordinating with more complex models, such as using temperature-pressure profile data from CAMEMBERT simulations of well-studied targets like K2-18b \citep{Montet_2015} and GJ 1214b (\citealp{2009Natur.462..891C,Kempton_2023}) as inputs, or initializing with and/or validating against reference data from CREME. It is also possible to conduct 1-D model exoMIPs independently of other models, but designed such that their outcomes could serve various purposes. For example, 1-D model results may offer new ideas for parameterizations to be incorporated into the more complex models, in an effort to advance improvements in model performance. 1-D model results can also be used to test the underlying assumptions built into a given class of model. The current plan to use PIE outputs to test MALBEC codes for scenarios that span planetary regimes (e.g., hot Jupiters to sub-Neptunes to temperate terrestrials) is an example of the latter approach. 

Not all the considerations that are likely to arise for comparisons using particular models, and for particular planetary targets of interest, can be foreseen. If the history of CMIP and PMIP provides a guide, the tools we use to advance the modeling of exoplanet environments are likely to change considerably in the next 10 years. This is especially true if the machine learning techniques currently being developed for modern Earth climate model parameterizations \citep{2023Natur.621E..45B,acp-23-523-2023} and analyses are transferable to the planetary science realm.

In the meantime, there is much to learn about how exoplanet models handle climates that are (sometimes wildly) different from modern Earth. In preparation for the models and observations of the next decade and beyond, CUISINES aims to encourage constructive assessments of model performance among exoMIP participants, in much the same way that CMIP and PMIP have promoted model evaluations for the Earth climate science community. CUISINES is not in a position to say which models are “right” and which are “wrong,” since there are currently almost no observations beyond spectra with which to validate model results. Instead, CUISINES is focused on understanding where the differences between models arise, and how those differences shape perceptions of what other worlds may be like; THAI and MALBEC \citep{Villanueva_2024} have already shown that such work is possible. In the case of Earth-like exoplanets, this may include impressions that a target world is uninhabitable when further analysis and observations would show that it is (i.e., a false negative); alternatively, a target may seem habitable initially, but is later found not to be (a false positive). This CUISINES framework offers a foundation for future exoplanet model intercomparisons as we move on in this exciting era of planetary discovery.

\section{Summary} \label{sec:5}

Understanding how exoplanet atmosphere, radiative transfer, and retrieval models may produce different results for the same experimental design is critical if these models are to be used as guides to analyzing and interpreting data gathered by JWST and future observation missions. Stand-alone modeling studies may introduce confusion when experiment parameters vary between two or more studies of the same target so that a direct comparison between model results cannot be made. These studies may not also fully describe model parameterizations or the “tweaks” needed to bring a simulation to a successful conclusion, hampering efforts by other researchers to reproduce published results. To help address these issues, CUISINES has developed a framework consisting of five principles that allows researchers to work collectively and openly within the context of specific exoplanet model intercomparison projects (exoMIPs). The community input provided at the exoMIP design stage helps to ensure that project topics reflect areas of broad interest and utility. The collaborative nature of each exoMIP ensures that participating modeling groups can learn more about their own model’s performance compared to a community benchmark, and discover new ways to increase the robustness of their model results. The early sharing of exoMIP-specific protocols enhances participation; and the archiving of clearly described and readily accessible experimental setup information and model output enhances the potential for reproducibility as well as use of the model data beyond the planetary science community. In addition to exoMIP efforts, CUISINES will support contributions of single-model benchmark studies to BASIL that permit the community to evaluate the skill of new and updated models, as they become available. While CUISINES-endorsed exoMIPs will need to follow the framework proposed here, all exoplanet modelers are encouraged to adopt as much of the CUISINES framework as possible, to facilitate the community-wide interactions that will help advance the exoplanet modeling field in the years to come.

\begin{acknowledgments} We thank the Steering Committee of the Nexus for Exoplanet System Science (NExSS) Research Coordination Network for their support of the CUISINES Working Group, and the BUFFET Workshop that got the inaugural CUISINES exoMIPs off to an auspicious start. CUISINES Co-Chefs T.J.F., L.E.S., and A.Y. acknowledge support from the GSFC Sellers Exoplanet Environments Collaboration (SEEC), which is funded in part by the NASA Planetary Science Division's Internal Scientist Funding Model. L.E.S. and K.T. acknowledge support provided by NASA Earth and Planetary Science Division Research Programs, through ISFM work package ROCKE-3D at The Goddard Institute for Space Studies. Financial support to R.D. was provided by the Natural Sciences and Engineering Research Council of Canada (NSERC; Discovery Grant RGPIN-2018-05929), the Canadian Space Agency (Grant 18FAVICB21), and the European Research Council (ERC; Consolidator Grant 771620). J.H.M. acknowledges funding from the NASA Habitable Worlds program under award 80NSSC20K0230. D.A.C. acknowledges financial support from the Max Planck Society. N.J.M. acknowledges support from a UKRI Future Leaders Fellowship [Grant MR/T040866/1], a Science and Technology Facilities Funding Council Nucleus Award [Grant ST/T000082/1], and the Leverhulme Trust through a research project grant [RPG-2020-82]. GC acknowledges the financial support of the SNSF (grant number: P500PT\_217840). We appreciate the discussions with participants in the THAI and BUFFET workshops, whose feedback helped to shape the CUISINES framework as well as the protocols for our inaugural exoMIPs, and the reviewers whose comments have helped to improve the manuscript.  
\end{acknowledgments}

\bibliography{CUISINES.bib}
\bibliographystyle{aasjournal}
\end{document}